\documentclass[conference]{IEEEtran}
\IEEEoverridecommandlockouts

\usepackage[english]{babel}
\usepackage{blindtext}
\usepackage{tikz}
\usepackage{amsmath}
\usepackage{amssymb}
\usepackage{tabu}
\usepackage{adjustbox}
\usepackage{filecontents}
\usepackage{xspace}
\usepackage{multirow}
\usepackage{multicol}
\usepackage{subcaption}
\usepackage{titlesec}
\usepackage[font=small]{caption} 

\titlespacing*{\section}
{0pt}{1.5ex}{2.0ex}

\renewcommand{\baselinestretch}{0.958}   
\usepackage{tikz}
\usepackage{amsmath}

\usepackage{multicol}
\usepackage{comment}
\usepackage{float}
\usepackage{csquotes}
\usepackage{textcomp}

\usepackage{filecontents}
\usepackage{xspace}
\usepackage{multirow}

\usepackage{multicol}
\usepackage{lipsum}
\usepackage{mwe}
\usepackage{subcaption}

\definecolor{darkerblue}{RGB}{0,63,202}
\definecolor{sabzseyedi}{RGB}{14,109,25}
\definecolor{LightGray}{RGB}{247,247,247}

\newcommand{\algo}{RecTen\xspace}  
\newcommand{\storyline}{StoryLine View\xspace} 
\newcommand{\tableview}{Table View\xspace} 

\newcommand{\reminder}[1]{{\textsf{\textcolor{red}  {[#1]}}}}

\usepackage{pifont} 

\usepackage{array} 
\newcolumntype{P}[1]{>{\centering\arraybackslash}p{#1}}

\usepackage{framed}

\definecolor{shadecolor}{RGB}{192,192,192}

\usepackage[linesnumbered,ruled,vlined]{algorithm2e}

\SetKwInput{KwInput}{Input}                
\SetKwInput{KwOutput}{Output}              

\begin{document}

\title{\algo: A Recursive Hierarchical Low Rank Tensor Factorization Method to Discover Hierarchical Patterns in Multi-modal Data\\
}

\author{
{\rm Risul Islam}\\
UC Riverside\\
risla002@ucr.edu
\and
{\rm Md Omar Faruk Rokon}\\
UC Riverside\\
mroko001@ucr.edu
\and
{\rm Evangelos E. Papalexakis}\\
UC Riverside\\
epapalex@cs.ucr.edu
 \and 
{\rm Michalis Faloutsos}\\ 
UC Riverside\\
michalis@cs.ucr.edu
} 

\maketitle

\newcommand{\risul}[1]{{\bf {\textcolor{green}{Risul:}}{\textcolor{blue}{#1}}}}

\newcommand{\rismi}[1]{{\bf {\textcolor{green}{R+M:}}{\textcolor{red}{#1}}}}

\newcommand{\miii}[1]{{\bf  {\textcolor{blue}{MF:}}{\textcolor{red}{#1}}}}

 \newcommand{\miok}[1]{{\bf {\textcolor{OliveGreen}{#1}}}}

\newcommand{\hide}[1]{}
\newcommand{\semihide}[1]{{\tiny #1}}
\newcommand{\changed}[1]{{\textcolor{blue}{#1}}}
\newcommand{\proofApp}[1]{{\emph{\bf Proof of #1.}}}
\newcommand{\mkclean}{
    \renewcommand{\reminder}[1]{}
    \renewcommand{\comment}[1]{}
    \renewcommand{\semihide}[1]{}
    \renewcommand{\changed}[1]{}
}

\newcommand{\RelThreads}{$R_t$}

\newcommand{\method}{{\tt HackerChatter}}
\newcommand{\source}{O}
\newcommand{\destination}{D}
\newcommand{\node}{\ensuremath{v}}
\newcommand{\edge}{\ensuremath{e}}
\newcommand{\graph}{\ensuremath{\mathcal{G}}}
\newcommand{\length}{\ensuremath{{l}}}
\newcommand{\prob}{\pi} 
\newcommand{\error}{e}
\newcommand{\fun}{f}
\newcommand{\loc}{\lambda}
\newcommand{\outdeg}{d_{out}}
\newcommand{\base}{b}
\newcommand{\dist}{\delta}
\newcommand{\ratio}{\rho}
\newcommand{\extrad}{\alpha}
\newcommand{\nodec}{\nu}
\newcommand{\pvalue}{p-value}
\newcommand{\mean}{average}
\newcommand{\std}{standard-deviation}
\newcommand{\ba}{\beta_{0}}
\newcommand{\bb}{\beta_{1}}
\newcommand{\bc}{\beta_{2}}

\newcommand{\HH}{Heavy-Hitters\xspace}
\newcommand{\CS}{Consumers\xspace}
\newcommand{\HP}{High Producers\xspace}
\newcommand{\AP}{Average Producers\xspace}
\newcommand{\LP}{Low Producers\xspace}
\newcommand{\PR}{\textit{Producers}\xspace}
\newcommand{\Rho}{\mathrm{P}}
\newcommand{\expnumber}[2]{{#1}\mathrm{e}{#2}}

\newtheorem{problem}{Problem}
\newtheorem{definition}{Definition}
\newtheorem{fact}{Fact}
\newtheorem{observation}{Observation}
\newtheorem{proposition}{Proposition}

\newcommand{\OffComm}{OffensiveCommunity\xspace}
\newcommand{\OffCommShort}{OffensComm.\xspace}
\newcommand{\HTS}{HackThisSite\xspace}
\newcommand{\WSForum}{WildersSecurity\xspace}
\newcommand{\Ash}{Ashiyane\xspace}
\newcommand{\WSShort}{WildersSec.\xspace}
\newcommand{\Ethical}{EthicalHackers\xspace}
\newcommand{\EthicalShort}{EthicHacks\xspace}
\newcommand{\darkode}{Darkode\xspace}

\newcommand{\TT}{{\em Hacks}\xspace}
\newcommand{\PS}{{\em Services}\xspace}
\newcommand{\AN}{{\em Alerts}\xspace}
\newcommand{\AG}{{\em Experiences}\xspace}

\newcommand{\MV}{{\em Malware - Virus}\xspace}
\newcommand{\HA}{{\em Hack - Account}\xspace}
\newcommand{\TG}{{\em Tutorial - Guide}\xspace}
\newcommand{\SB}{{\em Sell - Buy}\xspace}
\newcommand{\AV}{{\em Attack - Vulnerability}\xspace}
\newcommand{\VG}{{\em Video - Game}\xspace}
\newcommand{\CC}{{\em Credit Card}\xspace}

\newcommand{\seedMethod}{Initialization via domain adaptation\xspace}

\newcommand{\seedMethodAcro}{IDA\xspace}

\newcommand{\WordsFreq}{\textit{TextInfo}\xspace}
\newcommand{\IPP}{\textit{DecimalVal}\xspace}
\newcommand{\Mixed}{\textit{Mixed}\xspace}
\newcommand{\Cocluster}{\textit{Co-Cluster}\xspace}

\newcommand{\extUserWord}{\textit{ContextInfo}\xspace}
\newcommand{\extUserWordlong}{Contextual Information\xspace}

\newcommand{\PostTextlong}{Text information of the post\xspace}   
\newcommand{\postTextlong}{text information of the post\xspace}
\newcommand{\postText}{\textit{PostText}\xspace}

\newcommand{\mfal}[1]{{\bf{\textcolor{blue}{MF:}}{\textcolor{red}{#1}}}}
\newcommand{\mrem}[1]{{\bf{\textcolor{green}{\xspace#1}}}}

\newcommand{\idenp}{Identification\xspace}
\newcommand{\classp}{Characterization\xspace}

\newcommand{\crosstrain}{cross-training\xspace} 

\newcommand{\cseed}{Cross-Seeding\xspace} 

\newcommand{\cporting}{Basic\xspace} 

\newcommand{\kwords}{$W$\xspace}
\newcommand{\kwordsDef}{Word-Range\xspace}
\newcommand{\norm}[1]{\left\lVert#1\right\rVert}

\newcommand{\eatreminder}{\renewcommand{\reminder}{\hide}}

%

\begin{abstract}
How can we expand the tensor decomposition to reveal a hierarchical structure of the multi-modal data in a self-adaptive way? 
Current tensor decomposition provides only a single layer of clusters.
We argue that with the abundance of multi-modal data and time-evolving networks nowadays, the ability to identify emerging hierarchies is important. To this effect, we propose \algo, a 
recursive hierarchical soft clustering approach based on tensor decomposition. 
Our approach enables us to: (a) recursively decompose clusters identified in the previous step, and (b) identify the right conditions for terminating this process. 
In the absence of proper ground truth, we evaluate our approach with synthetic data and test 
its sensitivity to different parameters.
We also apply \algo on five real datasets which involve the activities of users in online discussion platforms, such as security forums. 
This analysis helps us reveal clusters of users with interesting behaviors, including but not limited to early detection of some real events like ransomware outbreaks, the emergence of a black-market of decryption tools, and romance scamming. To maximize the usefulness of our approach, we develop a tool which can help the data analysts and overall research community by identifying hierarchical structures. \algo is an unsupervised approach which can be used to take the pulse of the large multi-modal data and let the data discover its own hidden structures by itself.  

\end{abstract}

\section{Introduction}




{\em How can we identify hierarchical clusters in multi-modal data?} To answer this question, we leverage a well-established strategy, tensor decomposition. Tensor itself is a multi-dimensional representation of the data. Here, we call each dimension a ``mode". For instance, any tensor that captures the interaction of users with different threads at different times is a ``3-mode tensor" where the modes are the users, threads, and discretized times. Here, we use the terms {\it multi-modal} and {\it multi-aspect} interchangeably.   


Tensor decomposition has emerged as a  powerful
analytical tool with a rich variety of applications, but so far,
focuses more on identifying latent clusters and less on identifying hierarchical structure hiding in the data.
Current tensor-based approaches do an excellent job of identifying with hidden and inherent patterns in the form of clusters.
We argue that often behaviors and phenomena have an inherent hierarchical structure, which currently may be missed.

{\bf Problem:}
The main focus of this work is a relatively less explored question: 
how can we extend tensor decomposition to identify hierarchies
if such hierarchies exist in the data?
The problem we address here is as follows.
We are given a multi-dimensional dataset,  and we seek to  identify the potentially hierarchical structure present in the data. 
The key challenge here is that: (a) we do not know a priori anything about the data, such as the number of clusters or levels,
(b) we want to adapt to different levels of "sensitivity"
meaning that different parts of the data may hide more
layers of hierarchy than others.
Understanding the hidden structures, especially hierarchical structures, in the data is always a challenging task. 
We want to build on the power of Tensor decomposition by expanding it to address the above challenges. 

\begin{figure}[t]
    \centering
    \includegraphics[width=8.5cm, height=5.75cm]{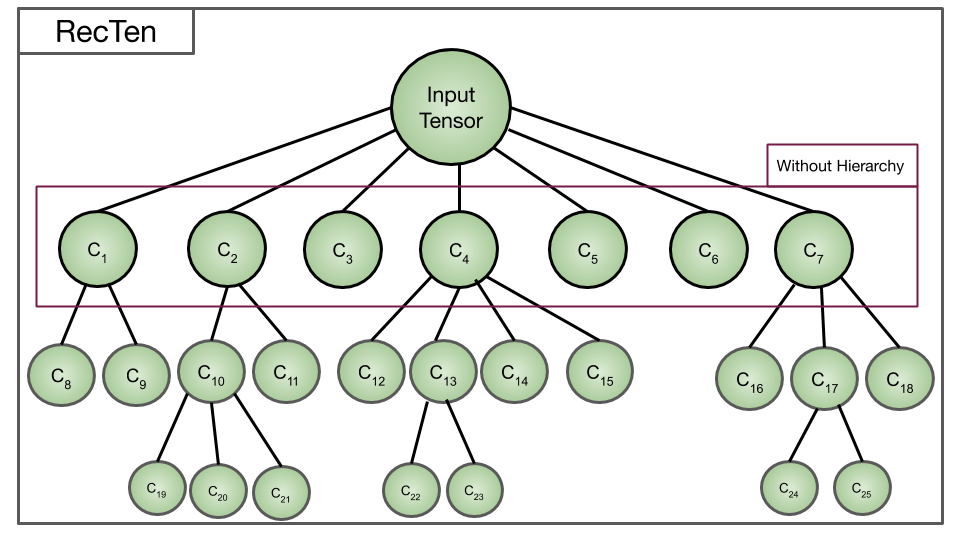}
    \caption{Output from \algo after applying it on ``Hack This Site" security forum data. Hierarchy up to three levels is presented here.}
    \label{fig:ourapproach}
\end{figure}


Formally, we state the problem as follows:

\begin{shaded*}
    Given a Tensor $T$, how can we expand the decomposition of $T$ in a recursive manner into a hierarchy of clusters, $C_i$? The input is a tensor and the output is a hierarchy of clusters. 
\end{shaded*}

The problem poses two main challenges:
(a) we want to recursively decompose clusters in each layer of the hierarchy, and (b) we want to identify the right conditions for terminating this process.


{\bf Related work:} 
Despite the vast literature on tensor decomposition, we are not aware of
any work that fully explores the possibility of hierarchical tensor decomposition.
We can group prior efforts into three main categories:
(a) discovering hierarchical structures using non-tensor approach,
(b) recursive hierarchical use of tensor decomposition \cite{abdali2020hijod, wang2015towards},
and
(c) other tensor decomposition approaches and applications.
We discuss these efforts in detail in  the {\em Related Work} Section.


{\bf Contribution:}
As our key contribution, we propose \algo, a  hierarchical soft clustering
approach based on tensor decomposition. 
Our approach provides the required mechanisms for recursively decomposing clusters, and  for terminating this recursive process. 
We  evaluate our proposed algorithm using both synthetic and real data.
We use the  synthetic data in order to evaluate the performance in the absence of proper ground truth. control the underlying structure of the dataset.  We also use this synthetic data to evaluate the sensitivity of three internal parameters, which enables us to provide recommendations for hands-free operation. In our evaluation, we find that \algo performs favorably when compared to other traditional and state-of-the-art methods.

We provide indirect evidence of the usefulness of \algo in isolating the data patterns by finding interesting clusters. These clusters correspond to online users with interesting behaviors.
We conduct a large scale study on 5 real datasets using \algo and explore the hidden hierarchical patterns within these datasets. 
For example, we analyze
three security forums: Offensive Community \textbf{(OC)}, Ethical Hacker \textbf{(EH)}, Hack This Site \textbf{(HTS)}. 
We also apply \algo on two other datasets: a MultiPlayer Game and Hacking Cheats ({\bf MPGH}) forum dataset, and a dataset of authors on GitHub. We identify interesting activities there as well including revenge hacking, romance scamming, and ransomware explosion as we discuss later in {\em Application Results and Observations} Section.

{\bf A usable platform.} As a tangible contribution, we develop a powerful user-friendly platform that will be useful to researchers, security analysts, and data specialists. The key advantages of \algo platform are as follows:
(a) it operates in an unsupervised way, 
(b) it is user-friendly by being 
both automatic  and customizable, and 
(c) it provides an interactive way of getting clusters in a hierarchical manner which helps to understand the inherent patter in the original input tensor.

\section{Datasets}
\label{sec:data}

To show the effectiveness of \algo, we apply it on a total of five datasets from our archive. The datasets include (i) three security forum datasets, (ii) one gaming forum dataset, and (iii) malware repositories and their authors from GitHub dataset. All these datasets are private except the gaming forum and we wish to make these public once the work is accepted.  We provide a brief description of these datasets below.

{\bf a. Security forum datasets. }
We utilize the data that we collect from three security forums: Offensive Community (OC), Hack This Site (HTS), and Ethical Hackers (EH) ~\cite{secforums}. All the forums are in English language. In these forums,
{\it users} initiate discussion {\it threads} 
in which other interested users can {\it post} to share their opinion. The discussions expand both ``white-hat" and ``black-hat" skills. The datasets of the security forums span 5 years from 2013 to 2017.
Each tuple in each of our datasets contains the following information: user ID, thread ID, post ID, time, and post content. 

{\bf b. Gaming forum dataset. }
We also apply \algo on another different type of online forum of larger size to see if our approach would work equally well. For this reason, we apply our method on an online gaming forum, Multi-Player Game Hacking Cheats (MPGH) \cite{secforums}. MPGH forum is one of the largest online gaming communities with millions of discussions regarding different insider tricks, cheats, strategy, and group formation for different online games. The dataset was collected for 2018 and contains 100K comments of 37K users~\cite{pastrana2018crimebb}. The format of each tuple is the same as security forum datasets. 

\begin{table}[t]
    \caption{Basic statistics of our datasets.}
    \footnotesize
    \centering
    \begin{tabular}{|P{0.27\linewidth}|P{0.08\linewidth}|P{0.20\linewidth}|P{0.1\linewidth}|P{0.09\linewidth}|}
         \hline
         \textbf{Dataset} & \textbf{Users} & \textbf{Threads/ Repositories} &  \textbf{Posts} & \textbf{Active Days}\\
         \hline
         Offensive Comm. & 5412 & 3214 & 23918 & 1239\\
         \hline
         Ethical Hacker &  5482 & 3290 & 22434 & 1175\\
         \hline
         Hack This Site & 2970 & 2740 & 20116 & 982 \\
         \hline
         MPGH & 37001 & 49343 & 100000 & 289 \\
         \hline
         GitHub & 7389 & 8644 & - & 2225 \\
         \hline
    \end{tabular}
    \label{tab:stat}
    \vspace{-4mm}
\end{table}

{\bf b. GitHub dataset. } GitHub platform 
enables the software developers to create software repositories in order to  store, share, and collaborate on projects and provides many social-network-type functions.

We  define some basic terminology here. 
We use the term {\it author} to describe a GitHub user who has created at least one repository.
A {\it malware repository} contains  malicious software and
a {\it malware author} owns at least one such repository.
Apart from creating a repository, users of GitHub 
can perform different {\it types of actions} including {\it forking, commenting}, and {\it contributing} to other repositories. {\it Forking} means creating a clone of another repository. A forked repository is sometimes merged back with the original parent repository, and we call this a {\it contribution}. Users can also {\it comment} by providing suggestions and feedback to other authors' repositories. Each tuple is represented in the format: malware author ID, malware repository ID, action type, time, and repository content.

We use a dataset of 7389 malware authors 
and their related 8644 malware repositories, which were identified by prior work~\cite{rokon2020source}.
This is arguably the largest malware archive of its kind with repositories spanning roughly 11 years.

The basic statistics of the datasets are shown in Table \ref{tab:stat}.

\section{Our Approach}
\label{sec:approach}

We present, \algo,  a novel tensor-based multi-step recursive approach that identifies patterns in an unsupervised way.  
Fig. \ref{fig:ourapproach} provides
the basic workflow of \algo. Algorithm 1 also provides the high level pseudo-code. The simplest idea is that: we decompose a tensor into clusters at level 1. Each cluster at level 1 will be considered as another tensor to be decomposed into clusters at level 2 and so on until some stopping criteria are met. We present the overview of  \algo in three steps: a)  Tensor-based clustering, b)  Prepare for the next level decomposition, and c) Termination conditions. We now discuss the basics of tensor decomposition and then the algorithmic steps below.


{\bf Tensors and decomposition.} A d-mode tensor~\cite{kolda2009tensor} is a d-way array (here $d\geq3$). So, we call \scalebox{0.85}{$I \times J \times K$} tensor a  ``3-mode'' tensor where ``modes'' are the number of dimensions to index the tensor; the ``modes'' can be $A={\{a_1, a_2, ..., a_I\}}, B={\{b_1, b_2, ..., b_J\}}$, and $C={\{c_1, c_2, ..., c_K\}}$. Each 3D {\it element/entity} of the tensor, \textit{X(i,j,k)}, captures the interaction of  $a_i$, $b_j$, and $c_k$ or zero in the absence of any interaction. 
In a decomposition, we  decompose a tensor into \textit{R} rank-one components, where \textit{R} is the rank of the tensor, as shown in Fig.~\ref{fig:2tensor}. That means, tensor is factorized into a sum of rank-one tensors i.e. a sum of outer products of three vectors (for three modes):  \scalebox{0.85}{$X \approx \sum_{r=1}^{r=R} A(:,r)\circ B(:,r)	\circ C(:,r)$} where \scalebox{0.85}{$A \in R^{I \times R}$ , $B \in R^{J \times R}$, $C \in R^{k \times R}$} and the outer product is derived by \scalebox{0.85}{$(A(:,r)	\circ B(:,r)$ $\circ C(:,r))(i,j,k)$ = $A(i,r)B(j,r)C(k,r)$} for all \textit{i, j, k}.
Each component represents a latent pattern in the data, and we refer to it as a {\bf cluster}.
For example, one such cluster in OC security forum  represents a group of 29 users that are active in the first weekends of July 2016 and discuss ``multi-factor authentication failure" in a group of 72 threads.
Each cluster is defined by three vectors, one for each dimension, which show the ``{\it Participation Strength}" of each element for that cluster.
Typically, one considers a threshold to filter out elements that do not exhibit significant {\it Participation Strength}, as we discuss later.

\subsection{Step 1: Tensor-based clustering} 


We provide a quick overview of the challenges  and algorithmic choices in our approach. 

\textit{a. What is the ideal number of components to target in the decomposition?} To answer this question, we use the AutoTen method~\cite{papalexakis2016automatic} and find the rank (R) of the tensor, which points to the ideal number of clusters to be decomposed into. AutoTen uses the {\it Core Consistency Diagnostic} metric in {\fontfamily{qcr}\selectfont CP\_ALS} and {\fontfamily{qcr}\selectfont CP\_APR} to find two probable ranks and finally chooses the max rank for the decomposition. So, the final rank, $R$, of a tensor is computed as follows: $R = max(R_{{\text{\fontfamily{qcr}\selectfont CP\_ALS}}}, R_{{\text{\fontfamily{qcr}\selectfont CP\_APR}}})$.

\textit{b. How can we decompose the tensor?}  We use the non-negative Canonical Polyadic or CANDECOMP/ PARAFAC (CP) decomposition to find the clusters.
\algo achieves this non-negative factorization by adding the non-negative constraint in CP decomposition.

\begin{figure}[t]
    \centering
    \includegraphics[ width=\linewidth]{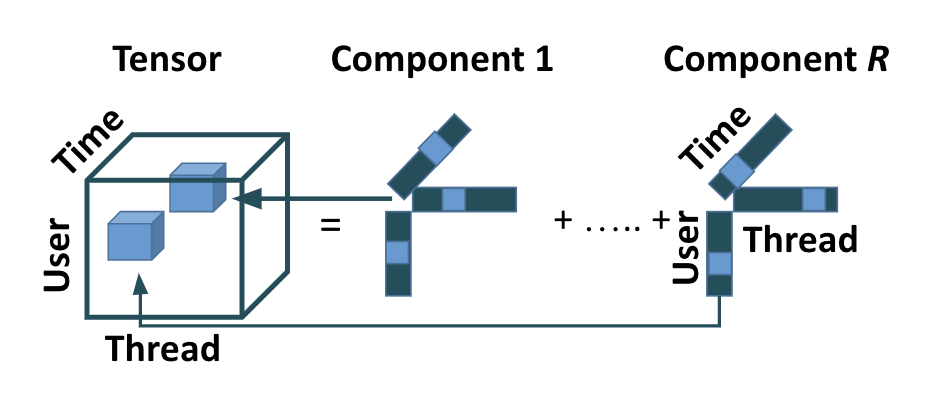}
    \caption{Visualization of tensor decomposition.}
    \label{fig:2tensor}
\end{figure}

\begin{figure}[t]
    \centering
    \includegraphics[ width=\linewidth, height=5.5cm]{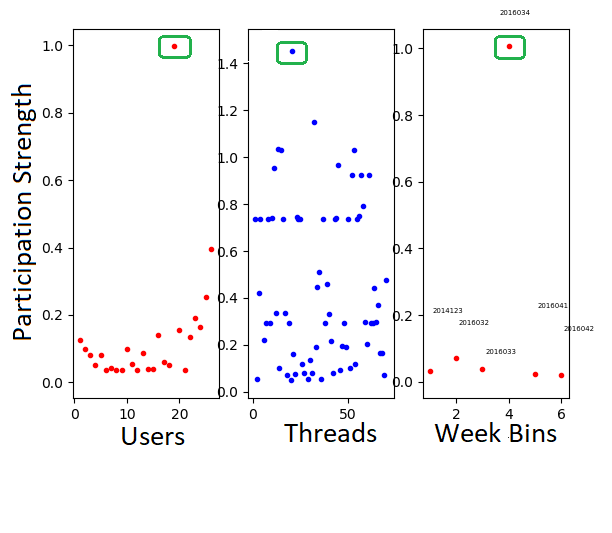}
    \caption{An example of a cluster (28 Users, 70 Threads, 6 Weeks) from OC. The intensity in each vector helps us identify users, threads and time intervals that are ``important" for the cluster. }
    \label{fig:cluster_example}
    
\end{figure}

\textit{c. How can we strike a balance on cluster size?}
Each cluster, derived from a 3D tensor, is defined by three vectors whose lengths are equal to the dimensions of the tensor as shown in Fig.~\ref{fig:2tensor}. However, \algo provides the functionality of having clusters with significant elements only. Therefore, we need a threshold to determine \enquote{significant participation in the cluster}, which is a common practice for (a) avoiding unreasonably dense clusters~\cite{sapienza2018non}, (b) enhancing interpretability, and (c) suppressing noise.
So, the challenge is to impose this sparsity constraint and eliminate the need for ad-hoc thresholding to find the clusters with only significant entities. 
Our solution is to add $L_1$ norm regularization with non-negative CP decomposition. $L_1$ regularization pushes the small non-zero values towards zero.
Therefore, for each vector, we filter out the zero-valued elements and produce clusters with significant elements only. In this way, we can eliminate the noisy entities having the least significant contributions in the cluster. The final model that we use for finding the clusters looks like this:

\scalebox{0.70}{%
$ \underset{A\geq 0, B\geq 0, C\geq 0}{\text{min}} \left\lVert X-D \right\rVert_{F}^{2} + \lambda ( \sum_{i,r} |A(i,r)| + \sum_{j,r} |B(j,r)| + \sum_{k,r} |C(k,r)|) $
}
where $\lambda$ is the sparsity regularizer penalty (set to 1) and \scalebox{0.75}{$D = \sum_{r} A(:,r)\circ B(:,r)\circ C(:,r)$}. To find the clusters, we solve the above equation. Since the equation is highly non-convex in nature, we use the well-established Alternating Least Squares (ALS) optimizer as the solver.  
An example of a cluster after filtering is shown in Fig.~\ref{fig:cluster_example}.


\subsection{Step 2: Processing for the next level decomposition}
\label{subsec:next_level_prep}
Having obtained the clusters at level 1 (in general at level $l$), our next goal is to decompose each cluster into further clusters if termination conditions are not met. 
The intuitive idea is to introduce some disturbance in the cluster so that the hidden patterns further gets isolated in separate clusters at the next level. That means, if there are multiple hidden structures/patterns are convoluted in a particular cluster, we enforce the cluster to be divided into further clusters. To achieve this goal, we face the following algorithmic challenges: 

{\bf Rank modification. } 
The main mathematical challenge in this phase is to answer the question: ``{\it How can we get the clusters ready to be decomposed further for the next level?" }  because the rank of every cluster is 1. That is why these clusters can not be decomposed directly. To decompose a cluster at level $l$ into multiple clusters we need to have the rank of the cluster at level $l$ greater than one. We propose to achieve this by introducing some disturbance in the cluster at level $l$. Recall that a cluster is also a tensor of rank one. Therefore, we  propose to introduce some disturbance by zeroing-out some carefully chosen subset of entities from this rank one tensor and make it ready to be factorized into further clusters because this zeroing-out strategy changes the rank of the cluster to $\geq 1$.

{\bf Choosing the ``target entities'' to zero-out.} The next question inherently comes into the picture is that ``{\it Which entities to choose for zeroing out? }". To zero out, we propose to select the non-zero valued entities stochastically but biased towards their values in the cluster. 

Let's assume 
$C_l$ be the rank-one cluster  at level $l$, $l \geq 1$. We get the processed Tensor $T_l$ with rank, $r>=1$, in the following manner:


We choose a subset of elements/entities stochastically biased towards their values, $C_l(i,j,k)$, and replace those with 0. We propose that the probability of an entity getting chosen to be replaced with 0 is inversely proportional to $C_l(i,j,k)$. That means, the higher the entity value, $C_l(i,j,k)$, the lower the probability of getting replaced with 0. Therefore, the probability of getting chosen for any non-zero point $p$ at $(i,j,k)$ for 3-mode Tensor, is formulated by the following formula:
$$P(\text{p is chosen}) = \frac{\frac{1}{C_l(p)}}{\sum_{q\in C_l} \frac{1}{C_l(q)}}$$  where $C_l(p) = C_l(i,j,k)$. After replacing the chosen entities with zero, we get the processed tensor, $T_l$, ready to be decomposed for the next level, $l+1$. In summary, we zero out some values from $C_l$ and get it ready for the next level decomposition.

{\bf Amount of deletion. } The next question arises is that: ``{\it What amount of entities should be zeroed-out?}" We propose to select $\epsilon\%$ of the total non-zero valued elements in $C_l$. We choose the next integer value after $\epsilon$ if we can not have exactly $\epsilon$. We find the suitable value of $\epsilon$ empirically described in {\it Evaluation of \algo} Section.

\subsection{Step 3: Termination condition}
\label{subsec:termination_condition}
We stop the recursive procedure of clustering when one of the two termination conditions is met:

{\bf a. Termination condition 1: Tensor size significantly small. } This preliminary termination condition suggests that when the cluster/tensor size is significantly small, we can not further break it. Now the obvious question is: ``{\it When do we call a cluster ``significantly small"?}" We say a cluster significantly small when the number of non-zero values in it
becomes less than {\it k\%} of the average number of non-zero values in the sibling clusters at that level. Let's say, at level $l$, tensor $T_{l-1}$ is factorized into a set of clusters $Y_l$, $x \in Y_l$ be a member cluster whose recursive factorization decision we are going to take,  $Z_l = Y_l \setminus x$ is the set of sibling clusters of $x$, $n(c)$ is the number of non-zero elements in cluster $c$, and $N(Z_L) = \frac{\sum_{z\in Z_l} n(z)}{|Z_L|} $ is the average number of non-zero elements in the sibling clusters, $Z_L$. Therefore, according to termination condition 1, we will not factorize cluster $x \in Y_l$ if:

 $$\frac{n(x) * 100 }{N(Z_l)} < k $$
 

{\bf b. Termination condition 2: AutoTen rank=1. }
Another termination condition of the recursive factorization  is that after preparing the cluster for the next level factorization, \algo will not factorize that particular cluster if its ideal rank provided by AutoTen is still 1. To summarize, if $R == 1$ for the cluster after processing, we skip the factorization of that particular cluster.

\begin{shaded*}
{\it Claim 1. Using \algo, zeroing-out of subset of non-zero entities from the rank-1 tensor changes the rank to $\geq 1$.} 
\end{shaded*}

We present the intuition behind proving this claim here. Lets say, we have a tensor $T$ and after deletion of some non-zero elements, we have the new tensor $\dot{T}$. Basically, we have to prove $rank(\dot{T}) - rank(T)>=0)$ given the situation.

Assume that, in \algo, zeroing-out of some non-zero entities in the rank 1 tensor leads to rank zero. But from the definition of rank-zero tensor, we know that rank-zero tensor has only the zero-valued elements, and thus introduces contradiction because it is not possible to have a zero-tensor in \algo. So, zeroing out can never lead to rank zero tensor, i.e. $rank(\dot{T}) - rank(T) \nless 0)$. 

Now, during zeroing out in \algo, we actually get a new tensor, $\dot{T}$, by zeroing-out non-trivial non-zero entries from $T$. Mathematically, this is done using element-wise Hadamard Product between the rank-one tensor, $T$, and a basis tensor $B$ i.e. $\dot{T} = T ./ B$ where $B$ is the basis tensor with elements either set (1) or reset (0). 
This type of zeroing-out is equivalent to introducing at least zero perturbation. Therefore, we can say $rank(\dot{T}) - rank(T)>=0$.

\begin{algorithm}[t]
\DontPrintSemicolon
  
  \KwInput{Root Tensor, T} 
  \KwOutput{Clusters arranged in hierarchical tree format}
  Clusters = [clusters from first level Decomposition of given root tensor, T]\\
  Final\_clusters\_tree.insert(T,null)\\
  Final\_clusters\_tree.insert(Clusters,T)\\
  \While{ Clusters != empty}
  {
        Temp\_clusters=[]\\
        \For{ each C in Clusters}
        {
            \If{Termination Condition 1}
            {
                continue
            }
            Construct $C_{new}$ by removing some entities from $C$ \\
            \If{Termination Condition 2}
            {
                continue
            }
            Decompose $C_{new}$ into new clusters $C_i$'s\\
            Temp\_clusters.insert($C_i$'s)\\
            Final\_clusters\_tree.insert($C_i$'s, C)\\
        }
        Clusters=Temp\_clusters\\
  }
    \Return Final\_clusters\_tree\\

\caption{
\begin{shaded*}
{\bf RenTen({\it T})} Algorithm to factorize a Tensor recursively to have hierarchical clusters.
\end{shaded*}
}

\label{algo:recten}
\end{algorithm}

\section{Evaluation of \algo}
\label{sec:evaluation}
In the absence of proper ground truth for evaluation, we create synthetic tensors and apply \algo on them. We  show the effectiveness of our method by applying it on five real datasets as well discussed at {\it Application Results and Observations} Section. However, since \algo is a multi-modal hierarchical soft clustering method and there is a lack of ground truth, it becomes inherently hard and challenging to perform certain tasks including (i) labeling the extracted clusters, (ii) finding suitable evaluation metrics, and (iii) finding suitable comparison methods. We discuss these challenges in the following subsections.

\begin{figure}[t]
    \centering
    \includegraphics[width=\linewidth,height=5cm]{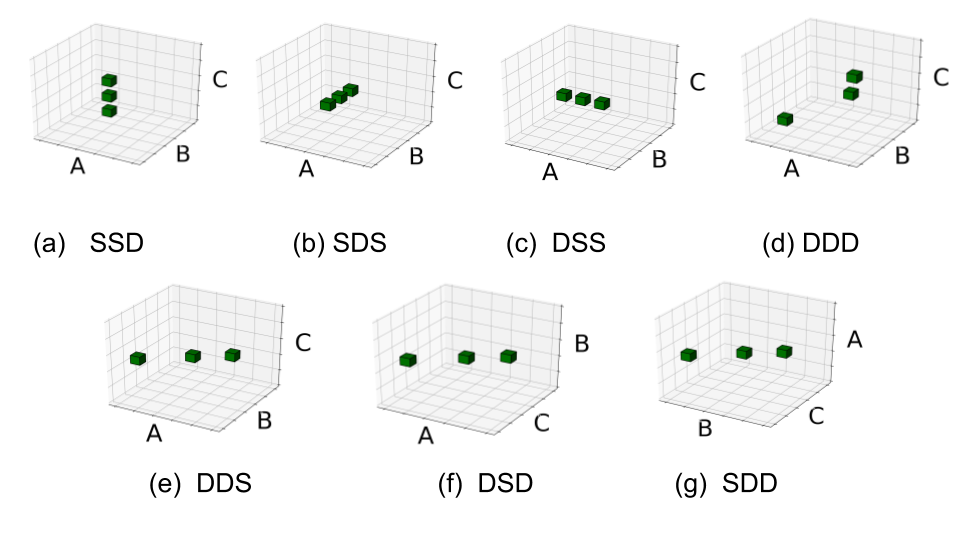}
    \caption{Injected non-hierarchical pattern/clusters of different Types in our Synthetic tensor.}
    \label{fig:flatsynthetic}
\end{figure}

\subsection{Synthetic Tensor Construction}
\label{subsec:synthetictensor}
To evaluate \algo, we construct synthetic tensors and demonstrate the effectiveness of \algo in finding both hierarchical and non-hierarchical clusters. We describe the injection of both non-hierarchical and hierarchical clusters in synthetic tensors below.

{\bf A. Injecting non-hierarchical clusters in Synthetic Tensor.}
We inject some non-hierarchical patterns in the form of clusters in a synthetic tensor demonstrated in Fig. \ref{fig:flatsynthetic}. We consider a 3-mode synthetic zero-tensor,$S$, with dimension $300\times 300 \times 30$ which is reasonable enough to emulate realistic patterns. Lets assume the modes are $A$, $B$, and $C$. $a_i, b_j, c_k$ are indices along modes $A$, $B$, and $C$. To imitate different types of overlapping and non-overlapping non-hierarchical clusters present in the data, we inject the following clusters in $S$ shown in Fig. \ref{fig:flatsynthetic}:

{\it a. SSD cluster pattern.} We insert three {\it SSD} cluster patterns, $P_1-P_3$, in $S$. These clusters have {\bf S}ame $a_i$'s, {\bf S}ame $b_j's$, but {\bf D}ifferent $c_k$'s. {\it b. SDS cluster pattern.} 3 {\it SDS} clusters, $P_4-P_6$, inserted; {\bf S}ame $a_i$'s, , {\bf D}ifferent $b_j$'s, {\bf S}ame $c_k's$. {\it c. DSS cluster pattern.} 3 clusters, $P_7-P_9$, inserted; {\bf D}ifferent $a_i$'s, {\bf S}ame $b_j$'s, {\bf S}ame $c_k's$. {\it d. DDD cluster pattern.} 3 clusters, $P_{10}-P_{12}$, inserted; {\bf D}ifferent $a_i$'s, {\bf D}ifferent $b_j$'s, {\bf D}ifferent $c_k's$. {\it e. DDS cluster pattern.} 3 clusters, $P_{13}-P_{15}$, inserted; {\bf D}ifferent $a_i$'s, {\bf D}ifferent $b_j$'s, {\bf S}ame $c_k's$. {\it f. DSD cluster pattern.} 3 clusters, $P_{16}-P_{18}$, inserted; {\bf D}ifferent $a_i$'s, {\bf S}ame $b_j$'s, {\bf D}ifferent $c_k's$. {\it g. SDD cluster pattern.} 3 clusters, $P_{19}-P_{21}$, inserted; {\bf S}ame $a_i$'s, {\bf D}ifferent $b_j$'s and {\bf D}ifferent $c_k$'s.

A total of 21 cluster patterns have been injected in $S$ which works as ground truth for the evaluation of \algo. However, to emulate the realistic patterns, we also follow some rules while injecting the cluster patterns. For instance, each pattern has a clustroid $\eta$. The patterns are formed with entities within $d$ Manhattan distance dispersion and $\rho$ concentration (entities/pattern). There are 8 overlapping patterns in $DDD$, $DDS$, $DSD$, and $SDD$ types. We regulated the pattern size and density by varying $d$ and $\rho$. The domain of each entity in each cluster pattern is a Gaussian distribution, $G(\mu=10,\sigma=3)$.

{\bf B. Injecting hierarchical cluster in synthetic tensor.} The goal here is to generate a time-evolving synthetic tensor with 3D realistic hierarchical patterns in it. We achieve this goal by following the steps below.

First, we inject 2D self-similar hierarchical patterns in the base slice (slice 0) of the synthetic tensor. Next, we emulate a time-varying tensor by making stack of the copies of the base slice but with substantial changes at each time slice. We follow the simplest but realistic approach borrowing the ideas from \cite{leskovec2010kronecker} and \cite{guigoures2012triclustering} to construct this synthetic tensor. 

\begin{figure}[t]
    \centering
    \includegraphics[ width=\linewidth, height=2.75cm]{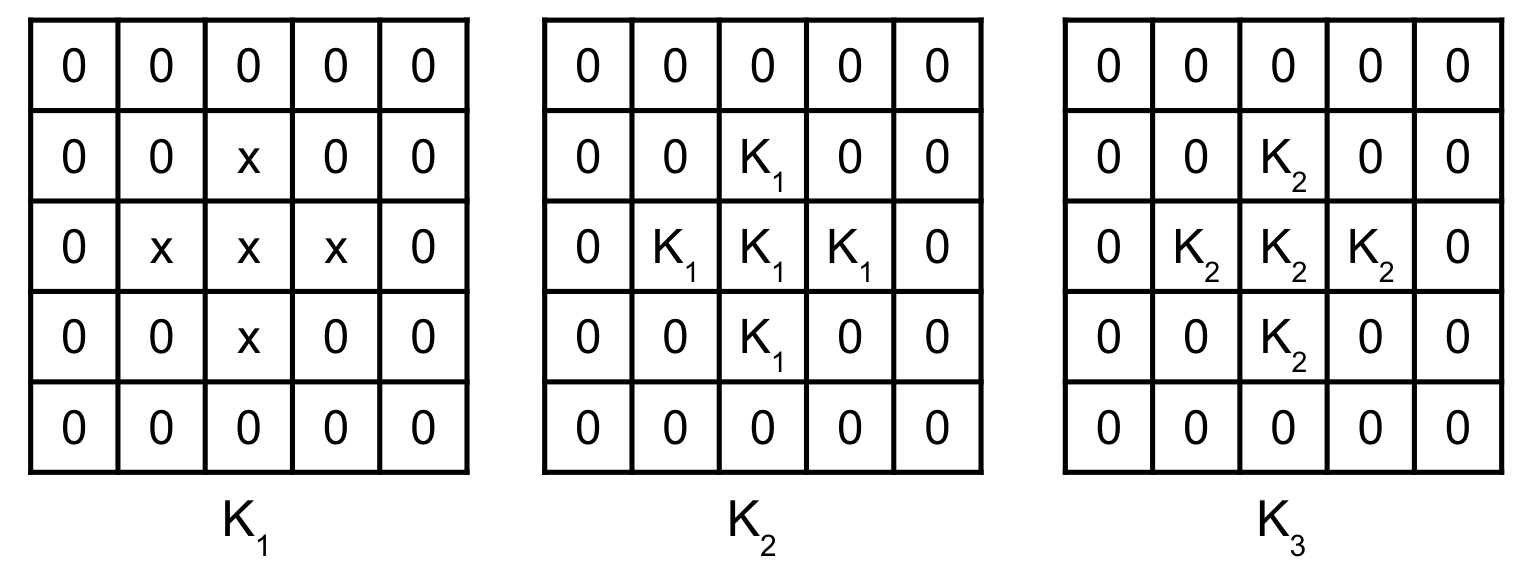}
    \caption{Example of Generation of Kronecker adjacency matrix $K_3$ for the base slice where $K_3 = K_2 \otimes K_1$ and $K_2 = K_1 \otimes K_1$. $\otimes$ is the Kronecker Multiplication operator.}
    \label{fig:kronecker}
\end{figure}

In more detail, we inject a hierarchical pattern utilizing Kronecker Multiplication of order three ($K_3$ adjacency matrix) \cite{leskovec2010kronecker} in the base slice (dimension 125X125) demonstrated in Fig. \ref{fig:kronecker}. The values of ``x'' in this Figure are drawn from a Gaussian distribution, $G(\mu=10,\sigma=3)$ and the white  places are all zeros. Next, we copy the base slice and make a stack of ten of these copies to emulate the time-evolving hierarchical pattern. To achieve substantial changes at each time slice, we randomly choose two zero-valued data points from each of the randomly chosen ten leaf clusters ($K_1$) and replace those with values from the above-mentioned Gaussian distribution. In this way, we can achieve a time-evolving synthetic tensor (dimension 125X125X10) with hierarchical structures in it as ground truth. In tree view, this synthetic tensor yields a total of 30 hierarchical clusters in two different levels (5 at level 1 and 25 more at level 2). These 30 clusters act as ground truth for evaluating \algo. We want to show that \algo is capable of identifying the clusters at their respective levels.  To show the effectiveness of \algo against noise, we also add positive Gaussian noise $G(\mu=0,\sigma=1)$ in randomly chosen $n\%$ of the entities at each time slice. We discuss the effect of the noise on the performance of \algo below in this Section. 

\subsection{Evaluation Metrics}
\label{subsec:eval_metrics}

Evaluating the clustering methods is always hard and challenging, especially hierarchical multi-aspect soft clustering. Previous methods \cite{luu2011approach, rand1971objective, zhang1989simple} suggested the use of performance metrics: Total Purity, Rand Index, and Tree Edit Distance in such case. We now briefly discuss these metrics below.

{\bf A. Total Purity. }We use the metric {\bf Total Purity}({\it TP}) to evaluate \algo \cite{luu2011approach}. It is measured on a scale of 0-1 where {\it TP=1} means perfect clustering and {\it TP=0} means no clustering at all. In our case, {\it TP=1} means no impurity i.e. each entity is in the cluster of similar kind.

We formally define a measure of clustering quality that is based on labeling information. Let L represents the set of labels (for non-hierarchical synthetic tensor $|L|=21$). Let $l(x)$ be the true label ($C_1-C_{21}$) of point $x$.
The label of a cluster, $C_i$, is the true label of the majority of points in that cluster, denoted by
$$l(C_i) = argmax_{b\in L}|{x \in C_i| l(x) = b}|$$
The impurity of a cluster is the proportion of points whose label differs from the labels of their cluster. 

{\it Definition 1. The total impurity ($TI$) of a clustering algorithm $A$, which yields a set cluster patterns $C$, is defined as:}

$$ TI(A) = \frac{1}{|C|}\sum_{i=1}^{|C|} \frac{\sum {x\in C_i| l(x) \neq l(C_i)}}{|C_i|}$$

In our case, $C$ is the set of cluster patterns yielded in the output of \algo. Thus, Total Purity of \algo is defined as:

$$TP(\algo)= 1 - TI(\algo)$$

{\bf B. Rand Index. } The Rand Index (RI) \cite{rand1971objective} computes a similarity measure between two clusterings by considering all pairs of samples and counting pairs that are assigned in the same or different clusters in the predicted and true clusterings. RI is ensured to have a value close to 0.0 for total random clustering and exactly 1.0 when the clusterings are identical. Given a set, S, of n elements and two clustering algorithms, X and Y, to compare, the formula to calculate RI is:

$$RI = \frac{a+b}{\binom{n}{2}}$$

where a is the number of pairs of elements in S that are in the same cluster in X and in the same cluster in Y. b is the number of pairs of elements in S that are in the different clusters in X and in the different clusters in Y. In our case, n denotes the total number of non-zero elements in the synthetic tensor. For ease of evaluation and comparison, we set the regularizer penalty term $\lambda$ to zero and we study the effect of $\lambda$ later in this Section.

{\bf C. Tree Edit Distance.} The hierarchical structures injected in our synthetic tensor can be viewed as a tree and the output of \algo can be viewed as the same as well. Therefore, to show that \algo does capture the same/similar tree structure, we evaluate \algo using another metric, Tree Edit Distance (TED) \cite{zhang1989simple}. TED between two labeled trees, $T_1$ and $T_2$, is the number of insertion, renaming and deletion operation needed to map exactly one tree onto another. The lower the TED, the better the clustering algorithm w.r.t. ground truth.

\subsection{Parameter Tuning }
\label{subsec:eval}
We set the values of different parameters of \algo empirically by applying it on the synthetic tensors. We utilize TP and RI  evaluation metrics to do this. The results are discussed below.  

\begin{figure}[t]
    \begin{subfigure}{0.48\columnwidth}
    \includegraphics[height=4cm,width=4.5cm]{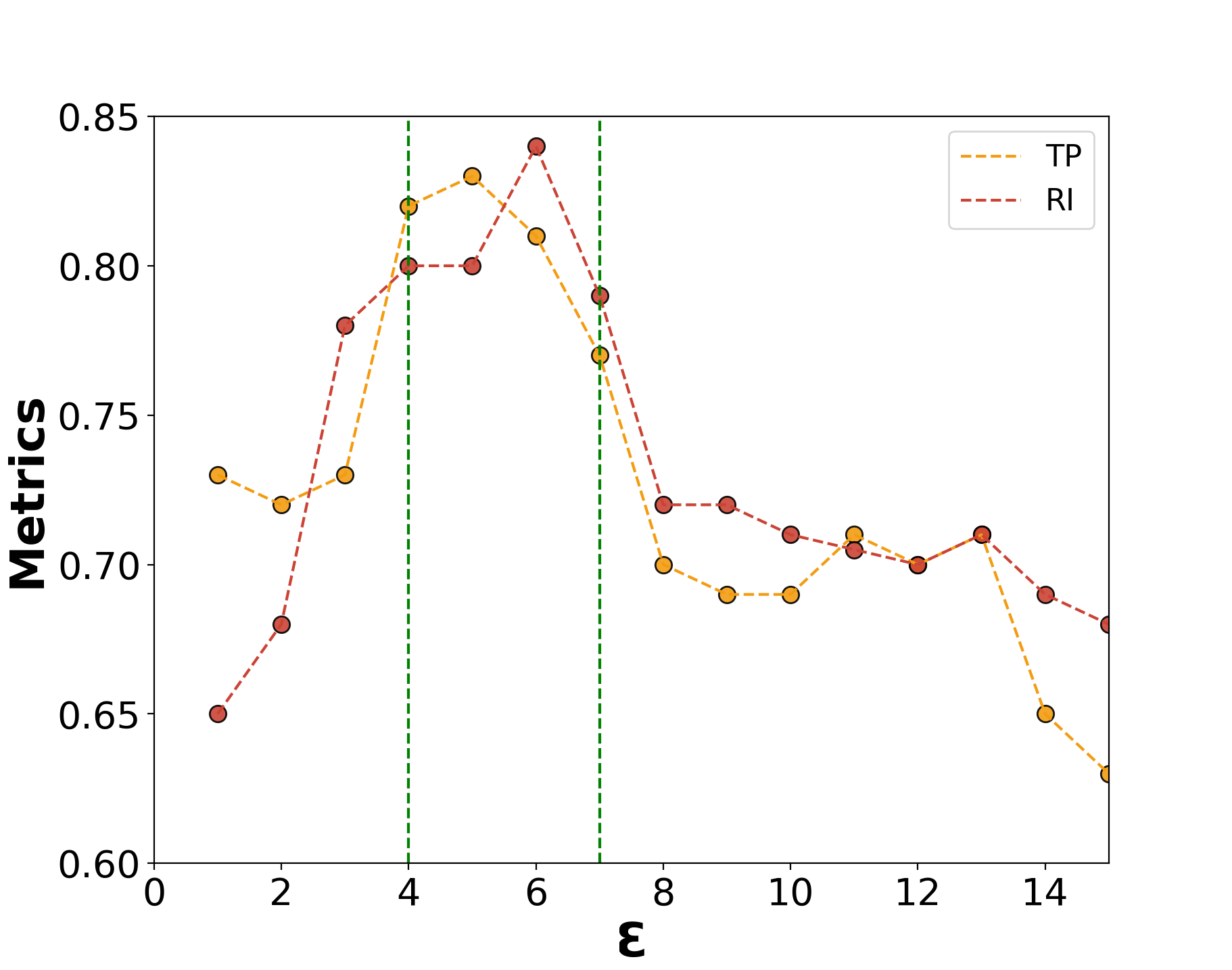}
        \caption{Non-hierarchical clustering.}
    \label{fig:evsmetricsflat}
    \end{subfigure}
    ~
    \begin{subfigure}{0.48\columnwidth}
    \includegraphics[height=4cm,width=4.5cm]{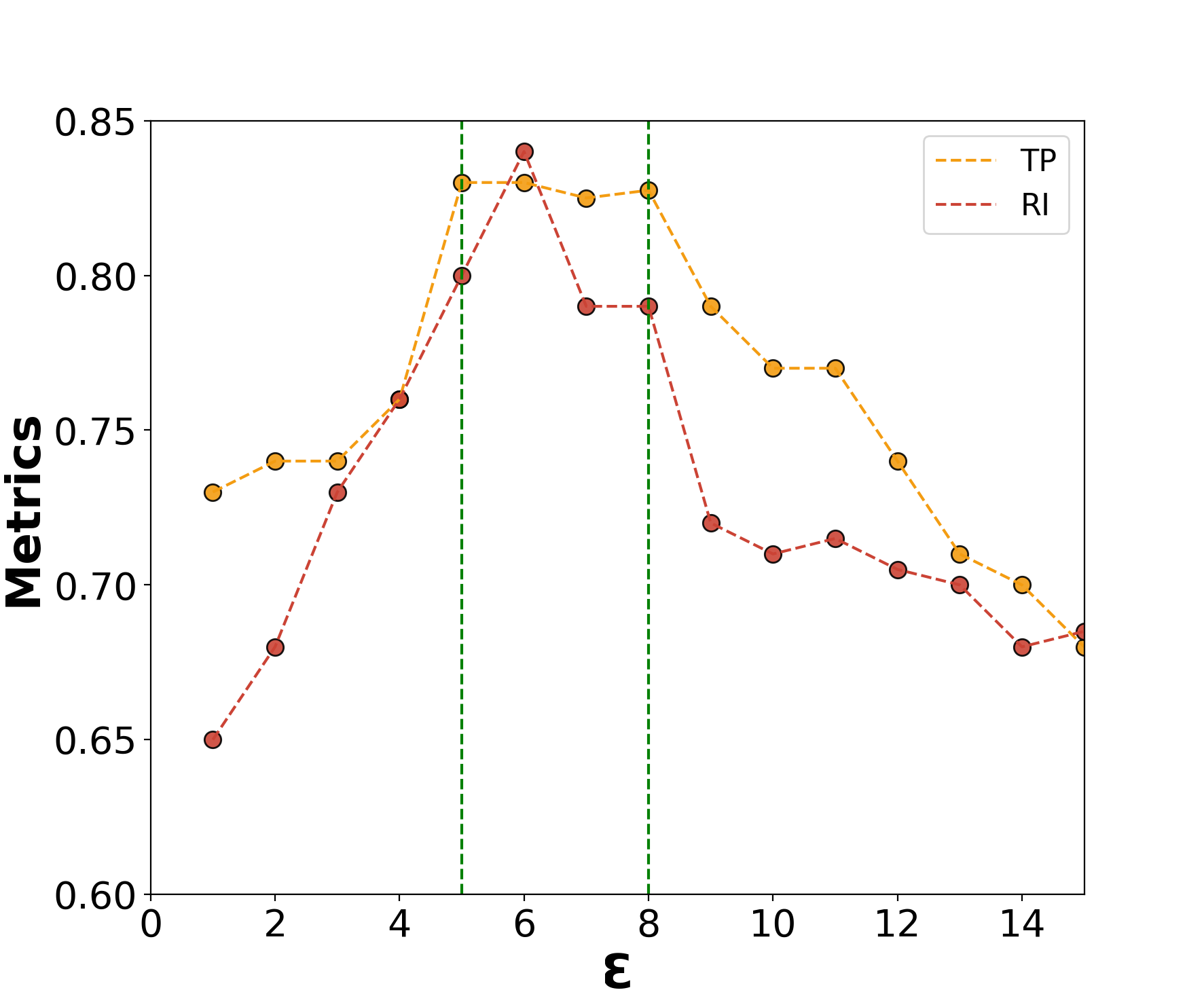}
        \caption{Hierarchical clustering.}
    \label{fig:evsmetricshierarchical}
    \end{subfigure}
    ~
    
    \caption{$\epsilon$ vs. TP and RI ($k=15$,$\lambda,n=0$). 
    }
    \label{fig:evsmetrics}
    \vspace{-2mm}
\end{figure}

{\bf a. Effect of $\epsilon$.} Choosing the right $\epsilon$ is crucial for \algo. Very small $\epsilon$ can yield no meaningful pattern whereas very large $\epsilon$ can eliminate significant amount of information from the tensor. So, the goal is to find a sweet region where we can unravel meaningful patterns with least amount of deletion. Fig. \ref{fig:evsmetrics} demonstrates that $\epsilon=[4,8]\%$ is our sweet region where we achieve maximum performance based on both metrics, TP and RI, for both non-hierarchical and hierarchical cluster extraction from synthetic tensors.

\begin{figure}[t]
    \begin{subfigure}{0.48\columnwidth}
    \includegraphics[height=4cm,width=4.5cm]{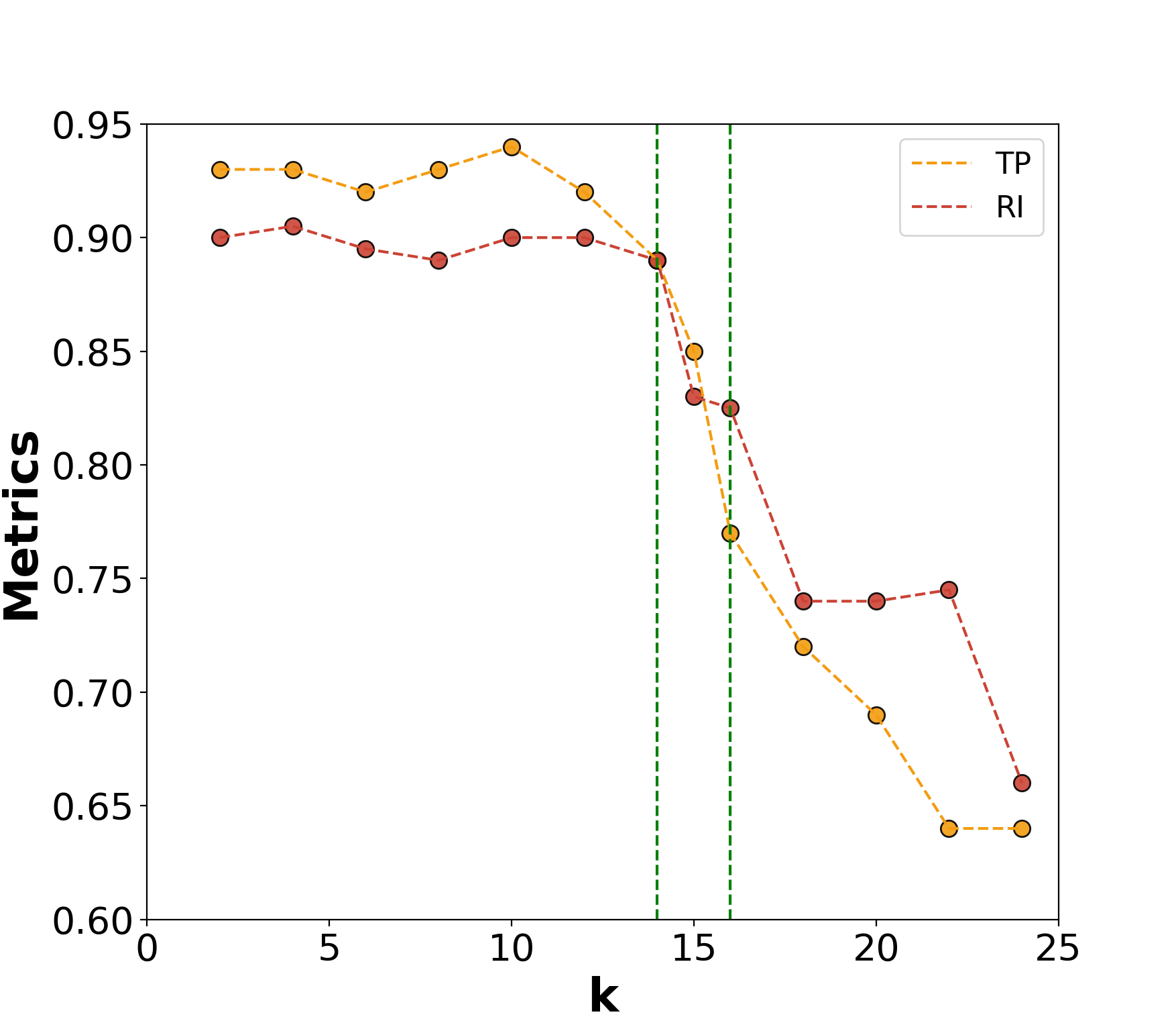}
        \caption{Non-hierarchical clustering.}
    \label{fig:kvsmetricsflat}
    \end{subfigure}
    ~
    \begin{subfigure}{0.48\columnwidth}
    \includegraphics[height=4cm,width=4.5cm]{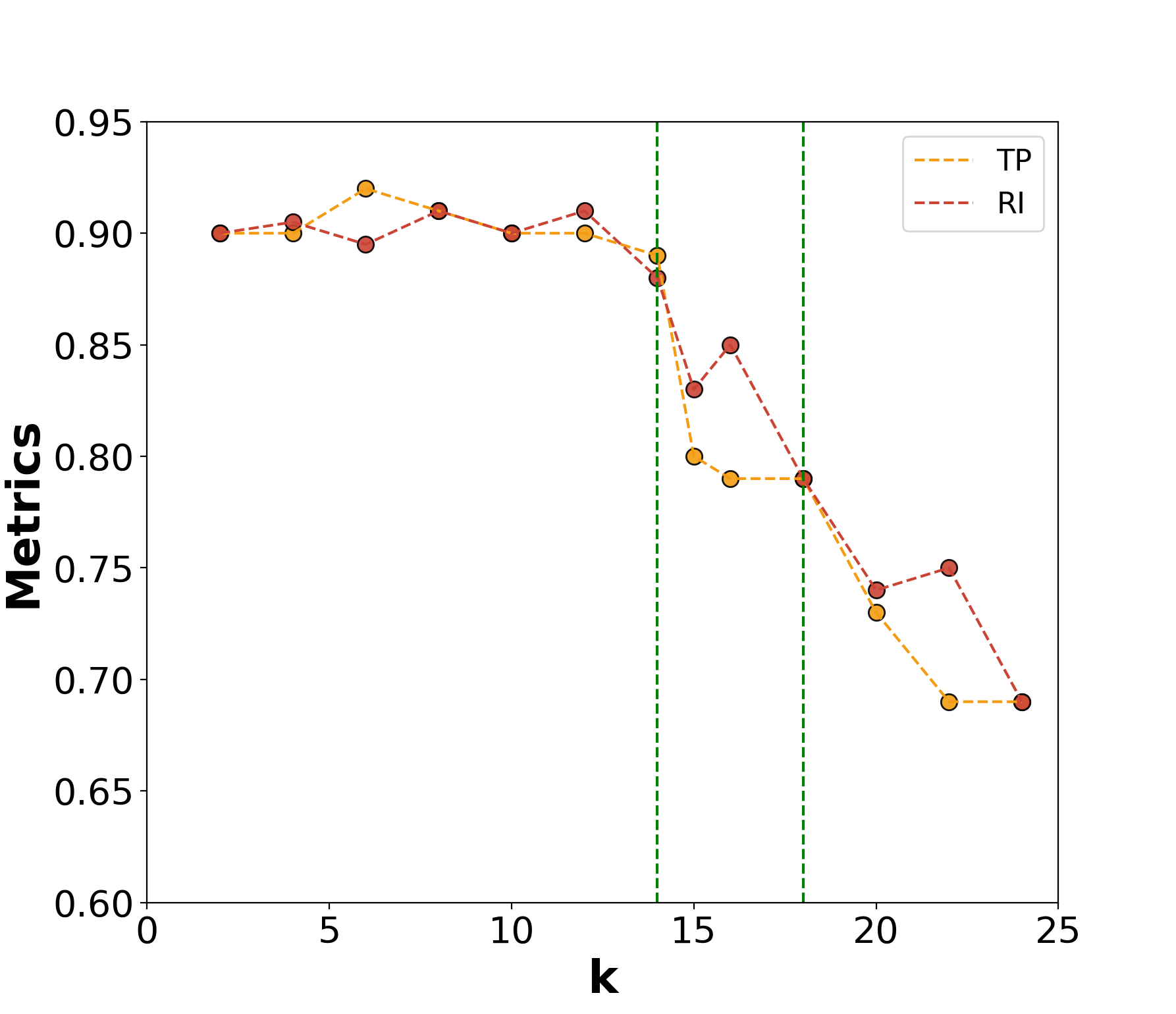}
        \caption{Hierarchical clustering.}
    \label{fig:kvsmetricshierarchical}
    \end{subfigure}
    ~
    
    \caption{$k$ vs. TP and RI ($\epsilon=6$,$\lambda,n=0$). 
    }
    \label{fig:kvsmetrics}
    \vspace{-2mm}
\end{figure}

{\bf b. Effect of $k$.}
Another crucial parameter of \algo is $k$ which determines when to stop our recursive factorization. Very low $k$ yields in small factorized clusters breaking down the pattern to even more parts (value of performance metrics close to 1) whereas very high value of $k$ will preserve multiple convoluted patterns in a single cluster (value of performance metrics far away from 1). Fig. \ref{fig:kvsmetrics} exhibits that, for our synthetic tensors, both hierarchical and non-hierarchical, $k=[14,18]\%$ is our sweet region where we achieve reasonably high performance from \algo based on both TP and RI performance metrics. Moreover, setting $k=15, \epsilon=6$, and $\lambda=0.8$, we find that \algo is able to find 20 clusters in level 1, and 3 clusters in level 2 from the non-hierarchical synthetic data. The algorithm would perform ideally if it could extract all 21 non-hierarchical clusters in level 1. Similarly, \algo captures 6 clusters at level 1, 25 clusters at level 2 and 1 clusters at level 3 whereas ideal performance would be identifying 5 clusters at level 1 and 25 clusters at level 2 from hierarchical synthetic data. This suggests the effectiveness of \algo in identifying hidden clusters from data. 

{\bf c. Effect of $\lambda$. }
\begin{figure}[t]
    \begin{subfigure}{0.48\columnwidth}
    \includegraphics[height=4cm,width=4.5cm]{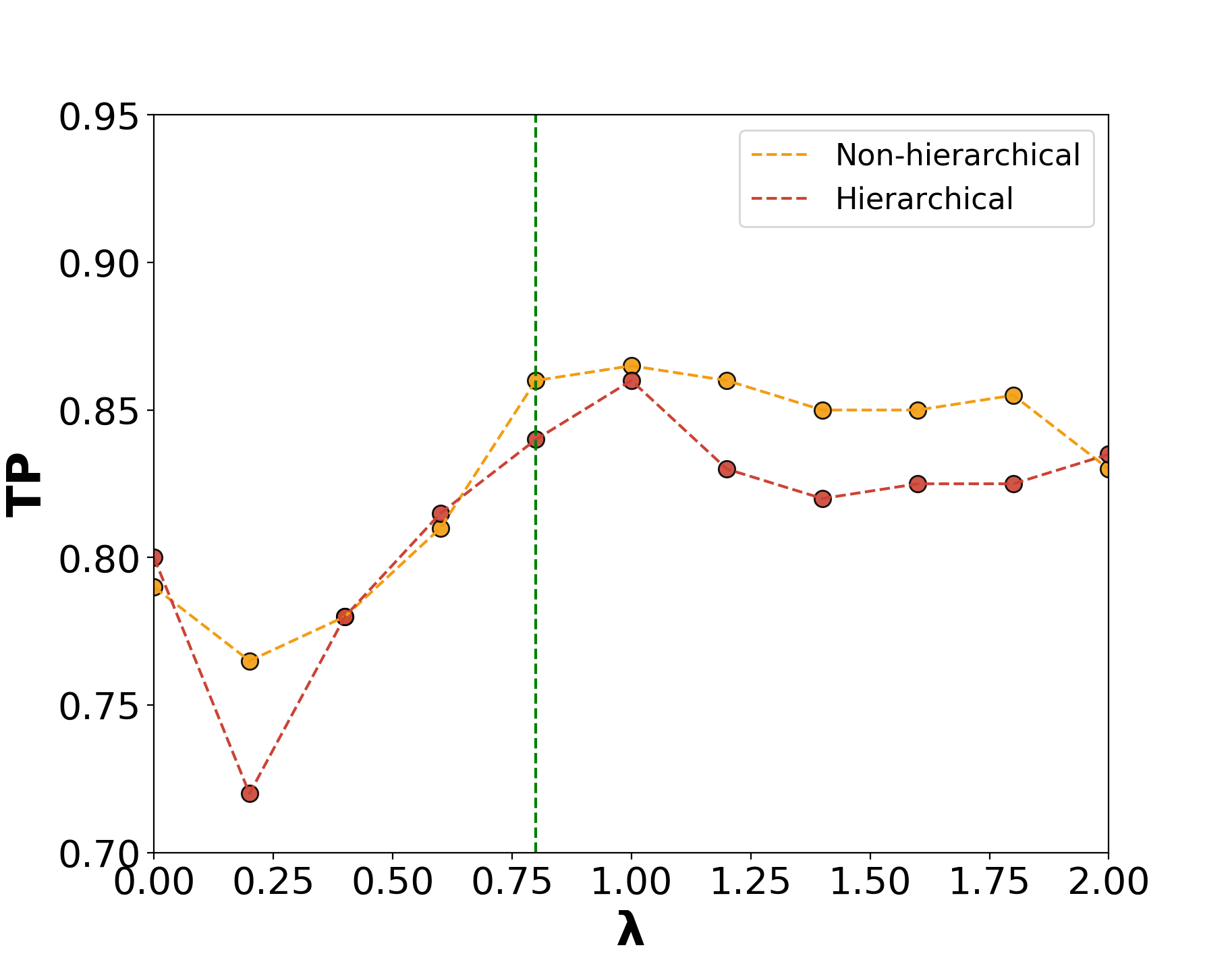}
        \caption{$\lambda$ vs. TP.}
    \label{fig:lambdavstp}
    \end{subfigure}
    ~
    \begin{subfigure}{0.48\columnwidth}
    \includegraphics[height=4cm,width=4.5cm]{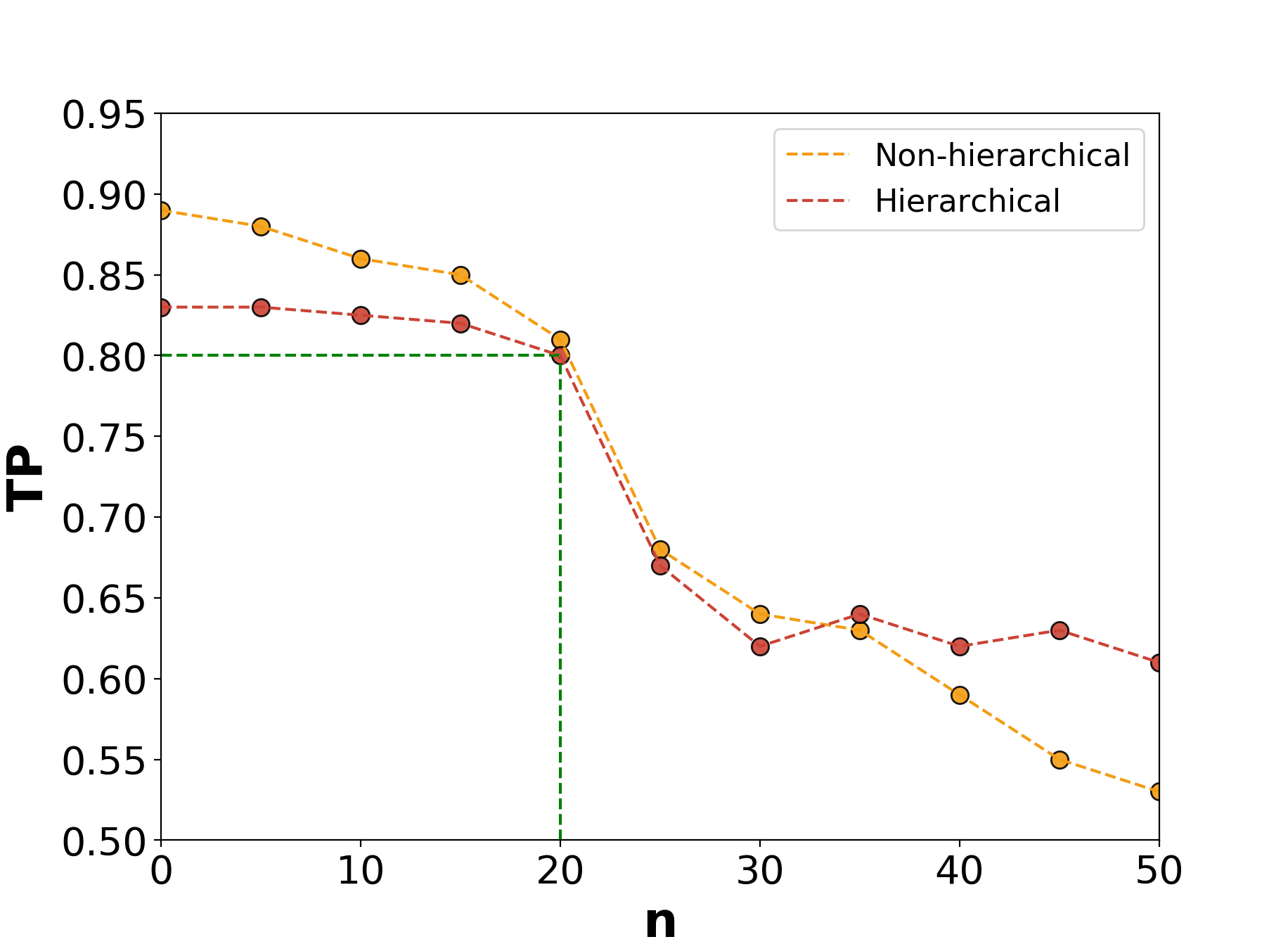}
        \caption{$n$ vs. TP.}
    \label{fig:nvstp}
    \end{subfigure}
    ~
    
    \caption{Effect of $\lambda$ and $n$ ($k=15$,$\epsilon=6$). 
    }
    \label{fig:lambdaandn}
    \vspace{-2mm}
\end{figure}
The sparsity regularizer penalty term, $\lambda$, is used to filter out the noisy data points while decomposition. If $\lambda=0$, then there will be no filtration at all. One key difference of \algo with other clustering algorithms is that \algo also offers clustering with only key elements by increasing $\lambda$. But very high value of $\lambda$ can yield clusters with high cohesiveness but minimal information. Fig. \ref{fig:lambdavstp} demonstrates the effect of $\lambda$ on the performance of \algo. Empirical result suggests that setting the value of $\lambda$ to 0.8 can yield maximum performance with reasonable information preserved.

{\bf d. Effect of noise percentage, $n$. } 
We analyze the performance of \algo in the presence of noise and find that \algo can offer reasonable performance for noisy data. Fig. \ref{fig:nvstp} demonstrates the performance of \algo in extracting patterns from both hierarchical and non-hierarchical synthetic tensors in the presence of noise. We observe a sharp drop ``knee" in Fig. \ref{fig:nvstp} which suggests that \algo can produce reasonable outputs for up to $n=20\%$ data noise.

{\bf e. Discussion on $d, \rho, \mu, \sigma$}
We also analyze the performance of \algo by varying dispersion $d$, concentration $\rho$, and data value distribution parameters ( $\mu$ and $\sigma$). We find that \algo offers good performance for different combinations of these parameters as well. We omit the results here due to space limitations. 







\subsection{Comparison with state-of-the-art methods}
\label{subsec:comparison}

To the best of our knowledge, we are the first to propose a method to extract hierarchical multi-modal clusters from multi-modal data using tensor decomposition. State-of-the-art methods find clusters from data represented in 2D matrix format whereas we find clusters from multi-modal data (tensor). Therefore, comparing results from two different kinds of data and cluster representation becomes inherently difficult. 

We try to compare the results with other widely-used and state-of-the-art methods anyway. To do this, we use four comparison methods including (i) widely-used basic Ward's method for Agglomerative Hierarchical Clustering Algorithm (AHC\_ward) \cite{ward1963hierarchical}, (ii) state-of-the-art frequency-based Agglomerative Hierarchical Clustering Algorithm (AHC\_freq) \cite{madheswaran2017improved}, (iii) local cores-based Hierarchical Clustering Algorithm (DLORE-DP) \cite{cheng2020dense}, and (iv) \algo without hierarchy (RecTen\_WH). We apply \algo as well as the above-mentioned Comparison Algorithms to extract structures from both the hierarchical and non-hierarchical synthetic tensors and compare their performance.

\begin{figure}[t]
    \begin{subfigure}{0.48\columnwidth}
    \includegraphics[height=4.5cm,width=4.5cm]{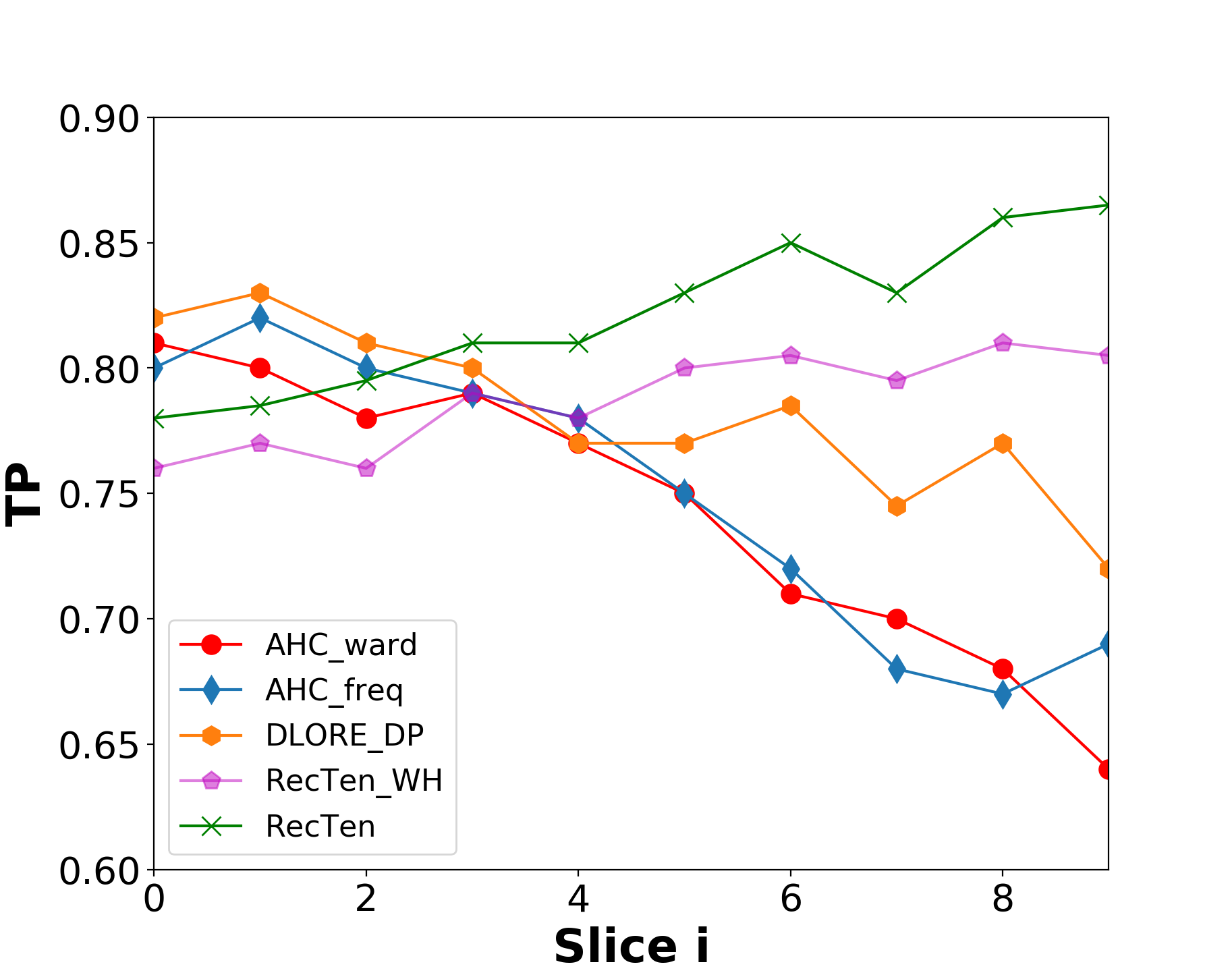}
        \caption{Performance comparison of \algo in terms of TP.}
    \label{fig:comparisontp}
    \end{subfigure}
    ~
    \begin{subfigure}{0.48\columnwidth}
    \includegraphics[height=4.5cm,width=4.5cm]{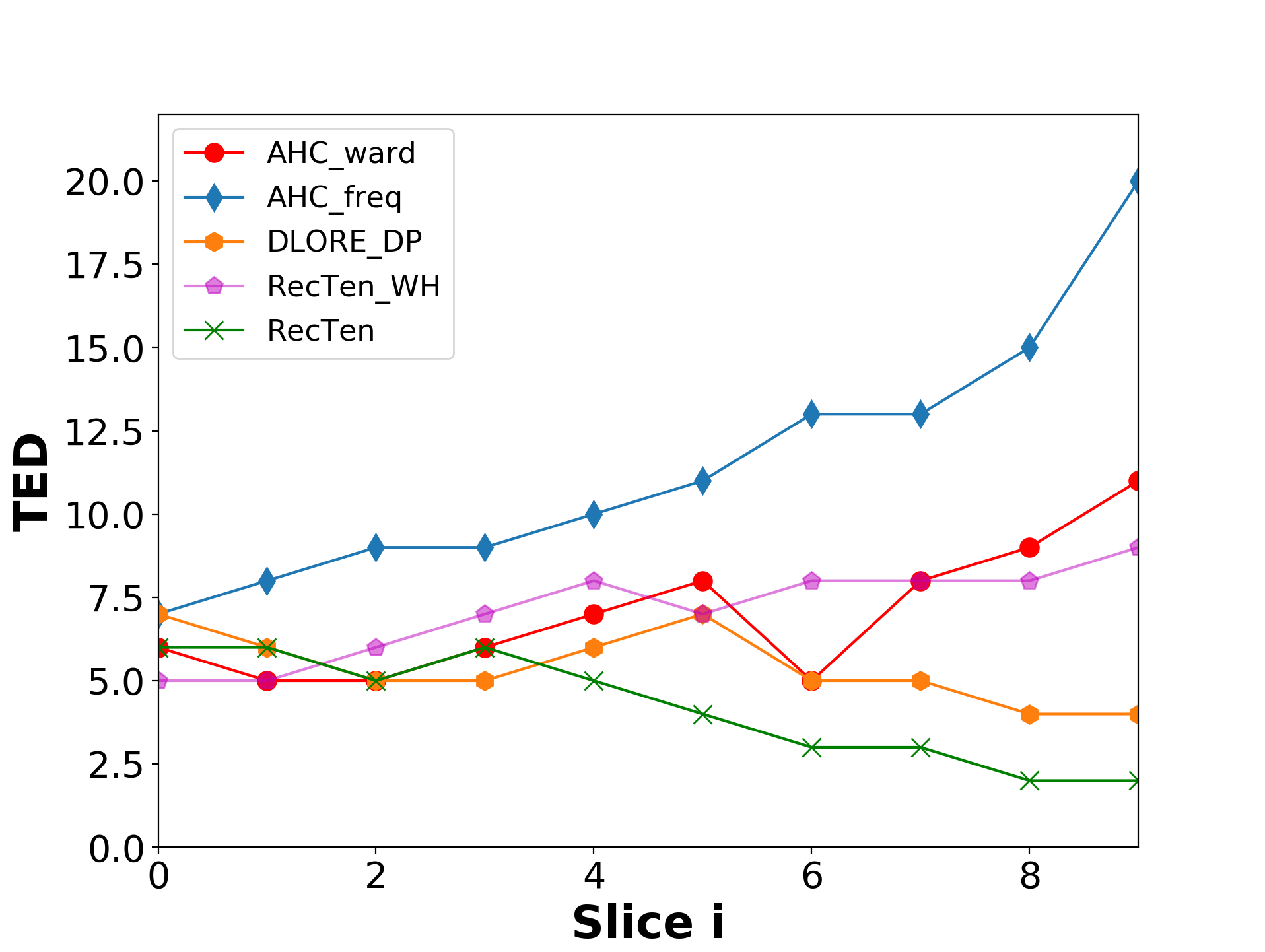}
        \caption{Performance comparison of \algo in terms of TED.}
    \label{fig:comparisonted}
    \end{subfigure}
    ~
    
    \caption{Performance comparison of \algo and our Comparison Algorithm for hierarchical cluster discovery. ($k=15$,$\epsilon=6,\lambda=0.8$). 
    }
    \label{fig:comparison}
    \vspace{-2mm}
\end{figure}

Now, as a desperate attempt to compare the Comparison Algorithms, applied on data represented in 2D matrix, with \algo, applied on 3D tensor, we follow a tricky way. We compare the performance results from Comparison Algorithms applied on 2D slice $i$ of the synthetic tensors with performance results from \algo applied on 3D sub-tensor constructed from slice 0 to $i$. The comparison results are depicted in Fig. \ref{fig:comparisontp} and \ref{fig:comparisonted}. We find that \algo underperforms than others except RecTen\_WH for the base slice $i=0$. DLORE-DP actually produces better result than any of the other methods up to slice 2. But  as $i$ increases more and new non-zero entities are added on the slices, \algo outperforms all the other Comparison Algorithms. Fig. \ref{fig:comparisonted} demonstrates the performance of \algo for hierarchical structures discovery compared to other methods in terms of TED performance metrics. Note that the lower the TED is, the better the performance is. Similar to hierarchical pattern discovery, \algo does better than the Comparison Algorithms in finding non-hierarchical clusters as well. We omit this similar result due to space limitations.

\section{Application Results and Observations}
\label{sec:results}

We provide proof of the effectiveness of \algo by finding interesting hierarchical clusters from five different real datasets, discussed in {\it Datasets} Section. The findings that we report here can help a security analyst mostly to a great deal. We discuss the findings below.

{\bf a. Results from security forums. } We construct a 3D tensor, $T$, by capturing the interaction of users with different threads at different weekly discretized times. Each element, $T(i,i,k)$ of the input tensor captures the interaction (in terms of the number of posts) of user $i$ with thread $j$ at  discretized week \textit{k} or zero in the absence of such interaction. We then fed the input tensor, $T$m to \algo. We find a total of 101 clusters from three security forums (41 from OC, 27 from HTS, and 33 from EH) arranged in hierarchical format (maximum depth 4). 

Upon investigating the cluster contents from OC forum, we find a cluster at level 1 revolving around Ransomware related discussion. The discussion started at Dec 2015 and further instigated in Feb 2016 (observed from the time dimension of the identified cluster) which mirrors the outbreak of SimpleLocker ransomware at that time. The detected cluster actually indicated the early warning of the coming ransomware disaster around the world. However, upon further investigation of that cluster by going one level down, we identified 12 threads containing 34 sellers of the decryption tools. By diving deep into the next level further, we identify a famous company, MDS, also selling decryption tools in this underground market.

Investigating one of the identified clusters at level 1 from HTS forum, we identify a bunch of threads by user {\it DoSman} offering free trial of his attack tools. Diving deep into the next level, we find posts related to selling DOS attack tools and phishing tools. EH forum also hides interesting clusters. For example, we identify {\it VandaDGod}, an expert Linux hacker, shared a popular tutorial series of hacking in Kali Linux in Nov 2017. Going one step down, we spot clusters related to `Hacking into Banks' and `Hacking Routers'. We find a series of posts to appoint new members for hacking into banks by {\it VandaDGod} in the next level. A simple internet search reveals that notorious hacker {\it VandaTheGod} is accused of hacking several government sites. This bears the indication that \algo does capture meaningful clusters.

{\bf b. Results from gaming forums. } Following the strategy similar  to security forums, we get a total of 233 clusters from MPGH organized in 6 levels (max depth of the output hierarchy tree is 6). We find clusters related to gaming strategies as well as some very peculiar and interesting clusters. For example, we identify a big cluster at level one revolving around different scamming and hacking related objections. Next level of this cluster comprises of clusters revolving around online gaming account scamming and `Romance Scamming'. It seems that in some online games where users can chat among themselves, scammers chat with female users to win their trust and scam for money. Also, we spot clusters related to searching for experienced hackers after a painful defeat in a game.  These surprising findings indicate that online gaming forums are being a new potential source of security threats.

{\bf c. Results from GitHub dataset. } Similar to security forums, we construct a 3D tensor for GitHub dataset as well. Each element, $T(i,i,k)$, of the input tensor captures the interaction (in terms of total number of create, fork, comment and contribution performed) of author $i$ with repository $j$ at  discretized week \textit{k} or zero in the absence of such interaction. Applying \algo on this tensor, we extract a total of 79 clusters in 4 levels (max depth 4). An interesting cluster contains Windows related malware in the first level and the next level contains Windows related ransomware. These clusters formed mainly in Jan 2016 when ransomware was spreading worldwide and malicious authors started developing more ransomware in GitHub inspired by the attack success. \algo can help the security enforcement authorities to keep track of which and how malware are being developed and getting popularity over time.

{\bf Empirical results comparison with TimeCrunch. } TimeCrunch \cite{shah2015timecrunch} is also a tensor based tool to discover patterns but it extracts only six fixed types of temporal structures, for example, oneshot full/near clique, periodic full/near bipartite cores etc. We apply TimeCrunch on three security forum datasets and find a total of only 17 structures. All these 17 structures are captured in \algo as well. \algo actually captures a total of 101 clusters from these security forums which is neither too few nor too many to analyze the clusterized actionable information. Therefore, \algo strikes a balance between finding too few and too many clusters to analyze. We omit the detailed finding of TimeCrunch due to space limitations.

{\bf Open-source for maximal impact. } Inspired by the success of \algo, we implement it in a tool format to help the research community. \algo tool can run with default parameter settings ($k=15$, $\epsilon=6, \lambda=0.8$) but any savvy end-user can optionally tune the parameters upon his/her needs and preferences. \algo provides clickable links on the output tree view by which the end-users can gauge through his/her cluster of interest and the contents of the clusters as well. 

{\bf Computational effort. } The computation required by \algo is not excessive. The average runtime for preparing the final hierarchical Tree view of the biggest forum with 100K posts, MPGH, takes only 2.39 minutes on average. 
Our experiments were conducted on a
machine with 2.3GHz Intel Core i5 processor and 16GB RAM.
We use Python v3.6.3 packages to implement all the modules of \algo. 
We believe that the runtime can be reduced to seconds if we use more powerful hardware.
These results suggest that \algo scales reasonably well in practice.
\section{Related Work}
\label{sec:related}


To the best of our knowledge, RecTen is the first of its kind to extract actionable hierarchical patterns from multi-aspect data. Our work is different from the following perspective: (i) it is unsupervised, (ii) it is applied on multi-aspect data, and (iii) it leverages tensor factorization in a recursive way. The related works can be divided into the following categories:

{\bf a. Discovering hierarchical structures using non tensor approach.}
Hierarchical structure discovery is a very common task in data mining. The algorithms that are being used mostly include but not limited to bottom-up Agglomerative Hierarchical Clustering, top-down Hierarchical k-means Clustering, and variations of these algorithms. The basic and widely used version of Agglomerative Hierarchical Clustering is Ward's method \cite{ward1963hierarchical}. Different variations of bottom-up Hierarchical Clustering are being used recently as well \cite{teffer2019adahash, mahalakshmi2020gibbs}. \cite{cheng2020dense} developed a local cores-based hierarchical algorithm for dataset with complex structures. Another work \cite{madheswaran2017improved} proposed an improved frequency-based agglomerative clustering algorithm for detecting distinct clusters on two-dimensional dataset. The application of these types of  bottom-up hierarchical extends but not limited to analyzing the sentiment of the people in different regions from  geo-location tagged twitter posts \cite{batarseh2018geo}, categorizing the  in-hospital fall reports of the patients \cite{liu2019exploring}, and document clustering \cite{lee2017hierarchical, zainol2017document}. Variations of Top-down hierarchical algorithm like hierarchical k-means is also being used in different domains ranging from modeling blast-produced vibration \cite{nguyen2019new} to large graph embedding \cite{nie2020unsupervised}. \cite{kuang2013fast} focuses on hierarchical document clustering leveraging non-negative matrix factorization. All the above-mentioned algorithms suffer from the same problem. These algorithms are applied on data represented in 2D matrix format. Therefore,  finding clusters in multi-modal data requires new strategy. The work of \cite{guigoures2012triclustering} focused on tri-clustering in time evolving graphs but did not utilize tensor factorization at all.





{\bf b. Recursive hierarchical use of tensors.} Although the recursive use of tensor is very rare, a very recent work of \cite{abdali2020hijod} introduces an embedding framework to detect fake news leveraging hierarchical tensor and matrix decomposition up to two levels. Though the authors call it a hierarchical model, it is completely different from our work because we use only tensor decomposition recursively. Another work \cite{wang2015towards} utilized recursive tensor decomposition for interactive topic hierarchy construction which is related but not actual recursive factorization like we do in this study.

\textbf{c. Other tensor decomposition approaches and applications.}
Tensor decomposition is a well-studied area with a wide range of diverse applications and domains ~\cite{kolda2009tensor}, ~\cite{liu2019community}, \cite{papalexakis2015understanding}, \cite{rettinger2012context}. \cite{shah2015timecrunch} proposed TimeCrunch which focused on mining some fixed temporal patterns from time-evolving dynamic graphs. Although the output of TimeCrunch are clusters like \algo, they are not organized in hierarchy.  

Relatively recently tensor-based techniques have been used in social media analytics as well to answer various questions. For example, ~\cite{papalexakis2015understanding} have used Tensor Decomposition to understand the multilingual social networks in online
immigrant communities. Another work of ~\cite{liu2019community} focused on community evolution by analyzing a feature matrix obtained by Tensor Decomposition. ~\cite{rettinger2012context} proposed a model for relationship prediction in social networks using context-aware Tensor Decomposition. Another type of tensor- based work includes community assignment of nodes in multi-aspect graph \cite{gujral2018smacd}.
We are not aware of any tensor-based multi-modal hierarchical pattern discovery algorithm. In our work, we adapted the CP decomposition \cite{kolda2009tensor, faber2003recent} to find the base clusters and then factorized the clusters recursively.






\section{Conclusion}

We propose, \algo,  an unsupervised-learning tensor-based approach to  systematically discover quality cluster patterns in a multi dimensional space. It has parameter-free settings as well as optionally parameter tuning capability.
We show the effectiveness of our approach by an extensive evaluation of the identified clusters from synthetic data as well as from five real datasets. 

\algo has three main advantages:
(a) it operates in an unsupervised way, 
(b) it generalizes well to both categorical and numerical multi-modal data and scales well for large dataset too,
and 
(c) in its tool version, it supports clickable Tree view to investigate the clusters further.

Our work is a step towards an automated unsupervised capability,
which can allow the data analysts and researchers to shift through the wealth of information that exists in massive multi-modal data.


%

%
\bibliographystyle{IEEEtran}
\bibliography{risul}

\begin{thebibliography}{10}
\providecommand{\url}[1]{#1}
\csname url@samestyle\endcsname
\providecommand{\newblock}{\relax}
\providecommand{\bibinfo}[2]{#2}
\providecommand{\BIBentrySTDinterwordspacing}{\spaceskip=0pt\relax}
\providecommand{\BIBentryALTinterwordstretchfactor}{4}
\providecommand{\BIBentryALTinterwordspacing}{\spaceskip=\fontdimen2\font plus
\BIBentryALTinterwordstretchfactor\fontdimen3\font minus
  \fontdimen4\font\relax}
\providecommand{\BIBforeignlanguage}[2]{{%
\expandafter\ifx\csname l@#1\endcsname\relax
\typeout{** WARNING: IEEEtran.bst: No hyphenation pattern has been}%
\typeout{** loaded for the language `#1'. Using the pattern for}%
\typeout{** the default language instead.}%
\else
\language=\csname l@#1\endcsname
\fi
#2}}
\providecommand{\BIBdecl}{\relax}
\BIBdecl

\bibitem{abdali2020hijod}
S.~Abdali, N.~Shah, and E.~E. Papalexakis, ``Hijod: Semi-supervised
  multi-aspect detection of misinformation using hierarchical joint
  decomposition,'' \emph{arXiv preprint arXiv:2005.04310}, 2020.

\bibitem{wang2015towards}
C.~Wang, X.~Liu, Y.~Song, and J.~Han, ``Towards interactive construction of
  topical hierarchy: A recursive tensor decomposition approach,'' in \emph{ACM
  SIGKDD}, 2015, pp. 1225--1234.

\bibitem{secforums}
{Online Forums}, ``{Ethical Hacker}, {Hack This Site}, {Offensive Community},
  {MPGH},'' \url{https://www.ethicalhacker.net/},
  \url{https://www.hackthissite.org/}, \url{http://offensivecommunity.net/},
  \url{https://mpgh.net/}.

\bibitem{pastrana2018crimebb}
S.~Pastrana, D.~R. Thomas, A.~Hutchings, and R.~Clayton, ``Crimebb: Enabling
  cybercrime research on underground forums at scale,'' in \emph{WWW}, 2018,
  pp. 1845--1854.

\bibitem{rokon2020source}
M.~O.~F. Rokon, R.~Islam, A.~Darki, E.~E. Papalexakis, and M.~Faloutsos,
  ``Sourcefinder: Finding malware source-code from publicly available
  repositories,'' in \emph{23rd International Symposium on Research in Attacks,
  Intrusions and Defenses}, In Press, 2020.

\bibitem{kolda2009tensor}
T.~G. Kolda and B.~W. Bader, ``Tensor decompositions and applications,''
  \emph{SIAM review}, vol.~51, no.~3, pp. 455--500, 2009, sIAM.

\bibitem{papalexakis2016automatic}
E.~E. Papalexakis, ``Automatic unsupervised tensor mining with quality
  assessment,'' in \emph{SDM16}.\hskip 1em plus 0.5em minus 0.4em\relax SIAM,
  2016, pp. 711--719.

\bibitem{sapienza2018non}
A.~Sapienza, A.~Bessi, and E.~Ferrara, ``Non-negative tensor factorization for
  human behavioral pattern mining in online games,'' \emph{Information},
  vol.~9, no.~3, p.~66, 2018, multidisciplinary Digital Publishing Institute.

\bibitem{leskovec2010kronecker}
J.~Leskovec, D.~Chakrabarti, J.~Kleinberg, C.~Faloutsos, and Z.~Ghahramani,
  ``Kronecker graphs: an approach to modeling networks.'' \emph{Journal of
  Machine Learning Research}, vol.~11, no.~2, 2010.

\bibitem{guigoures2012triclustering}
R.~Guigoures, M.~Boull{\'e}, and F.~Rossi, ``A triclustering approach for time
  evolving graphs,'' in \emph{2012 IEEE 12th International Conference on Data
  Mining Workshops}.\hskip 1em plus 0.5em minus 0.4em\relax IEEE, 2012, pp.
  115--122.

\bibitem{luu2011approach}
T.~Luu, ``Approach to evaluating clustering using classification labelled
  data,'' Master's thesis, University of Waterloo, 2011.

\bibitem{rand1971objective}
W.~M. Rand, ``Objective criteria for the evaluation of clustering methods,''
  \emph{Journal of the American Statistical association}, vol.~66, no. 336, pp.
  846--850, 1971.

\bibitem{zhang1989simple}
K.~Zhang and D.~Shasha, ``Simple fast algorithms for the editing distance
  between trees and related problems,'' \emph{SIAM journal on computing},
  vol.~18, no.~6, pp. 1245--1262, 1989.

\bibitem{ward1963hierarchical}
J.~H. Ward~Jr, ``Hierarchical grouping to optimize an objective function,''
  \emph{Journal of the American statistical association}, vol.~58, no. 301, pp.
  236--244, 1963.

\bibitem{madheswaran2017improved}
M.~Madheswaran \emph{et~al.}, ``An improved frequency based agglomerative
  clustering algorithm for detecting distinct clusters on two dimensional
  dataset,'' \emph{Journal of Engineering and Technology Research}, vol.~9,
  no.~4, pp. 30--41, 2017.

\bibitem{cheng2020dense}
D.~Cheng, S.~Zhang, and J.~Huang, ``Dense members of local cores-based density
  peaks clustering algorithm,'' \emph{Knowledge-Based Systems}, p. 105454,
  2020.

\bibitem{shah2015timecrunch}
N.~Shah, D.~Koutra, T.~Zou, B.~Gallagher, and C.~Faloutsos, ``Timecrunch:
  Interpretable dynamic graph summarization,'' in \emph{Proceedings of the 21th
  ACM SIGKDD International Conference on Knowledge Discovery and Data Mining},
  2015, pp. 1055--1064.

\bibitem{teffer2019adahash}
D.~Teffer, R.~Srinivasan, and J.~Ghosh, ``Adahash: hashing-based scalable,
  adaptive hierarchical clustering of streaming data on mapreduce frameworks,''
  \emph{International Journal of Data Science and Analytics}, vol.~8, no.~3,
  pp. 257--267, 2019.

\bibitem{mahalakshmi2020gibbs}
G.~Mahalakshmi, G.~MuthuSelvi, and S.~Sendhilkumar, ``Gibbs sampled
  hierarchical dirichlet mixture model based approach for clustering scientific
  articles,'' in \emph{Smart Computing Paradigms: New Progresses and
  Challenges}.\hskip 1em plus 0.5em minus 0.4em\relax Springer, 2020, pp.
  169--177.

\bibitem{batarseh2018geo}
F.~A. Batarseh, G.~Nambiar, G.~Gendron, and R.~Yang, ``Geo-enabled text
  analytics through sentiment scoring and hierarchical clustering,'' in
  \emph{2018 7th International Conference on Agro-geoinformatics
  (Agro-geoinformatics)}.\hskip 1em plus 0.5em minus 0.4em\relax IEEE, 2018,
  pp. 1--4.

\bibitem{liu2019exploring}
J.~Liu, Z.~Wong, K.-L. Tsui, H.-Y. So, and A.~Kwok, ``Exploring hidden
  in-hospital fall clusters from incident reports using text analytics.''
  \emph{Studies in health technology and informatics}, vol. 264, pp.
  1526--1527, 2019.

\bibitem{lee2017hierarchical}
C.-J. Lee, C.-C. Hsu, and D.-R. Chen, ``A hierarchical document clustering
  approach with frequent itemsets,'' \emph{International journal of engineering
  and technology}, vol.~9, no.~2, p. 174, 2017.

\bibitem{zainol2017document}
Z.~Zainol, S.~Marzukhi, P.~N. Nohuddin, W.~M. Noormaanshah, and O.~Zakaria,
  ``Document clustering in military explicit knowledge: A study on peacekeeping
  documents,'' in \emph{International Visual Informatics Conference}.\hskip 1em
  plus 0.5em minus 0.4em\relax Springer, 2017, pp. 175--184.

\bibitem{nguyen2019new}
H.~Nguyen, X.-N. Bui, Q.-H. Tran, and N.-L. Mai, ``A new soft computing model
  for estimating and controlling blast-produced ground vibration based on
  hierarchical k-means clustering and cubist algorithms,'' \emph{Applied Soft
  Computing}, vol.~77, pp. 376--386, 2019.

\bibitem{nie2020unsupervised}
F.~Nie, W.~Zhu, and X.~Li, ``Unsupervised large graph embedding based on
  balanced and hierarchical k-means,'' \emph{IEEE Transactions on Knowledge and
  Data Engineering}, 2020.

\bibitem{kuang2013fast}
D.~Kuang and H.~Park, ``Fast rank-2 nonnegative matrix factorization for
  hierarchical document clustering,'' in \emph{Proceedings of the 19th ACM
  SIGKDD international conference on Knowledge discovery and data mining},
  2013, pp. 739--747.

\bibitem{liu2019community}
Y.~Liu, G.~Yan, J.~Ye, and Z.~Li, ``Community evolution based on tensor
  decomposition,'' in \emph{ICPCSEE}.\hskip 1em plus 0.5em minus 0.4em\relax
  Springer, 2019, pp. 62--75.

\bibitem{papalexakis2015understanding}
E.~Papalexakis and A.~S. Do{\u{g}}ru{\"o}z, ``Understanding multilingual social
  networks in online immigrant communities,'' in \emph{WWW}, 2015, pp.
  865--870.

\bibitem{rettinger2012context}
A.~Rettinger, H.~Wermser, Y.~Huang, and V.~Tresp, ``Context-aware tensor
  decomposition for relation prediction in social networks,'' \emph{SNAM},
  vol.~2, no.~4, pp. 373--385, 2012, springer.

\bibitem{gujral2018smacd}
E.~Gujral and E.~E. Papalexakis, ``Smacd: Semi-supervised multi-aspect
  community detection,'' in \emph{ICDM}.\hskip 1em plus 0.5em minus 0.4em\relax
  SIAM, 2018, pp. 702--710.

\bibitem{faber2003recent}
N.~K.~M. Faber, R.~Bro, and P.~K. Hopke, ``Recent developments in
  candecomp/parafac algorithms: a critical review,'' \emph{Chemometrics and
  Intelligent Laboratory Sys}, vol.~65, no.~1, pp. 119--137, 2003, elsevier.

\end{thebibliography}

%

\end{document}